\documentclass[prl,twocolumn,preprintnumbers,amsmath,amssymb]{revtex4}

\usepackage{graphicx}
\usepackage{dcolumn}
\usepackage{bm}
\usepackage{color}
\usepackage{epstopdf}

\newcommand{\br}{ {\bm r}}
\newcommand{\bk}{ {\bm k}}

\begin{document}



\title{Rayleigh-Jeans condensation of classical light: \\
Observation and thermodynamic characterization}

\author{K. Baudin$^{1}$, A. Fusaro$^{1}$, K. Krupa$^{1,2}$, J. Garnier$^{3}$, S. Rica$^{4,5}$, G. Millot$^{1,6}$, A. Picozzi$^{1}$}
\affiliation{$^{1}$ Laboratoire Interdisciplinaire Carnot de Bourgogne, CNRS, Universit\'e Bourgogne Franche-Comt\'e, Dijon, France}
\affiliation{$^{2}$ Institute of Physical Chemistry Polish Academy of Sciences, Warsaw, Poland}
\affiliation{$^{3}$ CMAP, CNRS, Ecole Polytechnique, Institut Polytechnique de Paris, 91128 Palaiseau Cedex, France}
\affiliation{$^{4}$ University of Adolfo Ib\'a\~nez, Pe\~nalol\'en, Santiago, Chile}
\affiliation{$^{5}$ LadHyX, CNRS, Ecole Polytechnique, Institut Polytechnique de Paris, 91128 Palaiseau Cedex, France}
\affiliation{$^{6}$ Institut Universitaire de France (IUF), 1 rue Descartes, Paris, France}


\begin{abstract}
Theoretical studies on wave turbulence predict that a purely classical system of random waves can exhibit a process of condensation,
in analogy with the quantum Bose-Einstein condensation.
We report the experimental observation of the transition to condensation of classical optical waves propagating in a multimode fiber, i.e., in a conservative Hamiltonian system without thermal heat bath.
In contrast to conventional self-organization processes featured by the non-equilibrium formation of 
nonlinear coherent structures (solitons, vortices...), 
here the self-organization originates in the equilibrium Rayleigh-Jeans statistics of classical waves.
The experimental results show that the 
chemical potential reaches the 
lowest energy level at the transition to condensation, which leads to 
the macroscopic population of the fundamental mode of the optical fiber.
The near-field and far-field measurements of the condensate fraction across the transition to condensation are in quantitative agreement with the Rayleigh-Jeans theory.
The thermodynamics of classical wave condensation reveals that, in opposition to quantum Bose-Einstein condensation, the heat capacity takes a constant value in the condensed state and tends to vanish above the transition in the normal state.
Our experiments provide the  demonstration of a coherent phenomenon of self-organization that is exclusively driven by the statistical equilibrium properties of classical light waves.
\end{abstract}


\maketitle


Bose-Einstein condensation (BEC) has been reported in a variety of genuine quantum systems, such as ultracold atoms and molecules \cite{stringari}, exciton polaritons \cite{carusotto13} and photons \cite{weitz,fischer19}. 
On the other hand, several studies based on the wave turbulence theory \cite{zakharov92,newell01,nazarenko11,Newell_Rumpf,shrira_nazarenko13} predict that nonlinear waves can exhibit a phenomenon of condensation, in spite of the classical nature of the wave system \cite{newell01,nazarenko11,nazarenko05,PRL05,onorato06,berloff07,PD09,PRA11b,
laurie12,Fleischer,suret,PR14,nazarenko14,PRL18}.
Here we report the  observation of condensation that is driven by the equilibrium Rayleigh-Jeans (RJ) statistics of classical optical waves.

Various forms of condensation processes have been studied in the emerging key area of the quantum fluids of light \cite{carusotto13,faccio_lnp}.
In this context, condensation-like effects have been identified in optical cavity systems, which are inherently forced-dissipative systems operating {\it far from thermal equilibrium} \cite{PR14,conti08,fischer10,berloff13,turitsyn13,churkin15}.
On the other hand, equilibrium condensation mediated by the RJ distribution requires a (cavity-less) free propagation of the optical beam through a conservative (Hamiltonian) evolution.
However, as a consequence of the ultraviolet catastrophe inherent to classical waves, the RJ {\it condensation is not properly defined in free propagation}.
In this configuration, only a nonequilibrium transient process of condensation is experimentally accessible \cite{PRL18,chiocchetta16}.
This problem can be circumvented by considering a waveguide configuration, 
whose finite number of modes introduce an effective frequency cut-off that regularizes the RJ ultraviolet catastrophe \cite{PR14,christodoulides19}.
In this framework, a remarkable effect of spatial beam self-cleaning has been recently reported in multimode optical fibers (MMFs) \cite{wright16,krupa17}. 
Although recent works revealed that spatial beam cleaning is characterized by a transfer of power toward the low-order modes of the MMF, a detailed understanding of the mechanism underlying this effect is still debated \cite{krupa16,wright16,liu16,krupa17,conforti18,laegsgaard18,pod19,christodoulides19,PRL19,PRA19,
christodoulides19b}, see the review \cite{krupa19}.
Yet despite experimental progress, so far there is still no clear-cut demonstration of the phenomenon of RJ condensation of optical waves.

From a broader perspective,  wave condensation can be viewed as a self-organization process characterized by the formation of a {\it large scale coherent structure}, a universal behavior found 
in many fields of physics.
As a general rule, the formation of a coherent structure (e.g. solitons, vortices, shock waves...) {\it requires a strong nonlinear interaction regime} \cite{newell01,nazarenko11,Newell_Rumpf,shrira_nazarenko13,nazarenko05,PRL05,onorato06,
berloff07,PD09,PR14,rumpf01,zakharov04,rumpf_zakh09,janssen04,PRA11b,laurie12,nazarenko14,bagnato16},
a feature discussed through different recent experiments in water tanks  \cite{hassaini17}, vibrating elastic plates \cite{miquel13}, BECs \cite{bagnato16}, or nonlinear optics \cite{PRL18,pierangeli16,laurie12}.
More precisely, wave condensation is usually understood as an inverse turbulence cascade that increases the level of nonlinearity at large scales (i.e. low wave-numbers), up to a breaking point of the weak turbulence theory \cite{nazarenko11}.
Such a nonlinear stage of interaction is well-known in the focusing regime, where the (Benjamin-Feir) modulational instability leads to the generation of coherent soliton-like structures (`soliton condensation') \cite{nazarenko11,laurie12,rumpf01,zakharov04,rumpf_zakh09}. 
This fundamental nonlinear process is at the root of a variety of phenomena, e.g. optical filamentation and collapsing bursts \cite{newell01,nazarenko11}, the formation of droplets and bubbles in gravity waves \cite{nazarenko11}, or soliton-mediated supercontinuum generation in optics \cite{dudley06} and hydrodynamics \cite{chabchoub13}. 


\newpage

In contrast with this large variety of self-organization processes that occur {\it far from thermal equilibrium and require a strong nonlinear interaction}
\cite{newell01,nazarenko11,Newell_Rumpf,shrira_nazarenko13,nazarenko05,PRL05,onorato06,
berloff07,PD09,PR14,rumpf01,zakharov04,rumpf_zakh09,janssen04,laurie12,nazarenko14,bagnato16,
hassaini17,miquel13,pierangeli16},  
we report in our experiments a different mechanism of spontaneous formation of a coherent structure (condensate) that is driven by the {\it equilibrium RJ statistics in the weakly-nonlinear regime}.
Condensation originates in the RJ distribution for the following reasons: 
(i) The phase transition takes place when the chemical potential reaches the fundamental mode eigenvalue,  which leads to the macroscopic population of the fundamental mode of the MMF;
(ii) The condensate fraction across the transition
is in quantitative agreement with the {\it RJ equilibrium theory}; 
(iii) The nonlinearity is perturbative with respect to linear propagation, even in the strongly condensed regime.
Furthermore, the thermodynamics of classical condensation is characterized through 
the specific heat revealing noteworthy distinguished features in comparison with the quantum BEC transition.

Aside from its fundamental importance, light cooling and condensation find natural applications to achieve an accurate control of the coherence properties of optical beams in 
high-power multimode fibre sources \cite{wright17,christodoulides19}, which also attract a renewed interest for telecommunication applications \cite{kaminow13}.

{\it Experimental setup.-}
We study the 
spatial evolution of a speckle beam that propagates through a MMF featured by a parabolic-shaped index of refraction supporting $M \simeq 120$ modes.
The 2D parabolic potential $V(\br)=q |\br|^2$ 
is truncated at $V_0=q R^2$, where $R=26 \mu$m is the fiber radius and $q$ a constant determined by the fiber characteristics.
The eigenvalues are well approximated by the ideal harmonic potential $\beta_p=\beta_0(p_x+p_y+1)$, 
where $\{ p \}$ labels the two integers $(p_x,p_y)$  that specify a mode.
The truncation of the potential $V(\br) \le V_0$ and the corresponding finite number of modes $M$ introduce an effective frequency cut-off in the far-field spectrum $k_c=\sqrt{2V_0/\beta_0}/r_0$, where $r_0$ is the radius of the fundamental mode of the MMF \cite{supplemental}.

The source is a Nd:YAG laser ($\lambda_0=1.06 \mu$m) delivering sub-nanosecond pulses that are passed through a diffuser before injection into the MMF.
After propagation through a fiber length $L=12$m, both the near-field (NF) and the far-field (FF) intensity patterns are recorded with a (CCD) camera.
The NF intensity pattern $I_{\rm NF}(\br)=|\psi|^2(\br)$ provides a measurement of the power $N=\int I_{\rm NF}(\br) d\br$ and of the potential energy $E_{\rm pot}=\int V(\br) |\psi|^2(\br) d\br$.
The kinetic energy $E_{\rm kin}=\alpha \int |\nabla \psi(\br)|^2 d\br$ is retrieved from the FF pattern $I_{\rm FF}(\bk)=|{\tilde \psi}(\bk)|^2$, where $\alpha=1/(2 n_{\rm co} k_0)$ with $k_0=2\pi/\lambda_0$ the laser wave-number and $n_{\rm co}$ the refractive index. 
This provides the measurement of the linear contribution to the energy (Hamiltonian) $E=E_{pot}+E_{kin}$.
Projecting on the basis of the fiber modes, the power and energy read
$N=\sum_p n_p$, 
$E=\sum_p \beta_p n_p$,
where $n_p$ 
denotes the amount of power in the mode $\{p\}$ \cite{PR14} (here and below the sum $\sum_p$ is carried over the set of $M$ modes indexed by $\{p\}$).
We have confirmed by direct experimental measurements that the power $N$ and the energy $E$ are conserved during the propagation in the MMF \cite{supplemental}.


{\it Weakly nonlinear regime.-}
The speckle beam that propagates through the MMF exhibits fluctuations that vary over a linear propagation length $L_{lin}$ much smaller than the nonlinear length $L_{nl}$:
\begin{eqnarray}
L_{lin} \sim \beta_0^{-1} \sim 0.2{\rm mm} \ \ll \ L_{nl}=1/(\gamma N) \sim 0.3{\rm m},
\label{eq:Llin_Lnl}
\end{eqnarray}
where $\gamma$ is the nonlinear coefficient of the MMF.
The weakly-nonlinear regime (\ref{eq:Llin_Lnl}) is equivalent to say $\lambda_c \ll \xi$, where $\xi=\sqrt{\alpha L_{nl}} \simeq 130 \mu$m is the healing length, and $\lambda_c$ the transverse correlation length of the speckle beam, which is typically smaller 
than the fundamental mode 
of the MMF, $\lambda_c \lesssim r_0 =\sqrt{2\alpha/\beta_0}\simeq 4.7\mu$m~$\ll \xi$.
Such a large separation between linear and nonlinear scales  {\it prevents a process of nonlinear self-organization}, such as `soliton condensation' mediated by the 
Benjamin-Feir 
modulation instability (filamentation) of the speckle pattern.
In other words, the fact that the nonlinearity of the MMF is {\it focusing} ($\gamma >0$) does not play any role:
A spatial soliton cannot be generated since its width $\xi \gtrsim 130\mu$m  would be much larger than the radius of the MMF ($R=26\mu$m).

{\it Near- and far-field measurements of $(n_0^{eq},\mu)$.-}
In contrast with other experiments (e.g. \cite{weitz}), here no thermal bath is present: 
The thermalization to the RJ equilibrium is driven by the conservative Kerr nonlinearity 
during the propagation through the fiber.
As predicted by the wave turbulence kinetic theory and the numerical simulations \cite{PRA19}, thermalization occurs in a `closed' (Hamiltonian) system, 
where the {\it conserved energy $E$ is the control parameter} of the transition to condensation.
We have typically $L\sim 40 L_{nl} (\sim 60 \, 000 L_{lin})$ in the experiment, which is sufficient to achieve thermalization and condensation.
At equilibrium the modal populations follow the RJ distribution $n_p^{eq}=T/(\beta_p-\mu)$, $T$ and $\mu$ being the temperature and chemical potential \cite{PRA19,christodoulides19}. 
Accordingly we have $N=T\sum_p (\beta_p-\mu)^{-1}$ and $E=T \sum_p \beta_p /(\beta_p-\mu)$: 
The solutions to these equations show that $(\mu,T)$ are uniquely determined by $(N,E)$.
At variance with previous experiments of spatial beam cleaning \cite{wright16,krupa17,krupa16,liu16,pod19,krupa19}, here we study the transition to condensation in analogy with BECs, i.e. by decreasing the energy $E$ (`temperature') while {\it keeping constant the power} $N$ (`number of particles').
In our experiments $E$ is varied by passing the beam through a diffuser before injection in the MMF \cite{supplemental}.

The parabolic shaped potential $V(\br)$ allowed us to retrieve the condensate fraction $n_0^{eq}/N$ and the chemical potential $\mu$ from either the NF or FF intensity distributions at the output of the fiber.
Using the property that the Hermite-Gauss modes 
are invariant under Fourier transform, we follow the usual treatment in BECs and split the condensate and the incoherent contributions $N=n_0^{eq}+\sum_{p \neq 0}n_p^{eq}$, which gives 
\cite{supplemental}:
\begin{eqnarray}
&I_{\rm NF}^{\rm cond}(r)=n_0^{eq} r_0^2  w_0^2(r/r_0), \;  
I_{\rm FF}^{\rm cond}(k)=n_0^{eq} r_0^{-2} w_0^2(r_0 k), \quad
\label{eq:I_NF_fit_cond}
\end{eqnarray}
for the fundamental mode, and 
\begin{eqnarray}
I_{\rm NF}^{\rm incoh}(r)&=&\frac{(N-n_0^{eq})r_0^{-2}}{\sum_{p\neq 0} (\beta_p-\mu)^{-1}} \sum_{p\neq0} \frac{w_p^2(\br/r_0)}{\beta_p-\mu}, 
\label{eq:I_NF_fit_inc}\\
I_{\rm FF}^{\rm incoh}(k)&=&\frac{(N-n_0^{eq}) r_0^2}{\sum_{p\neq 0} (\beta_p-\mu)^{-1}} \sum_{p\neq0} \frac{w_p^2(r_0 \bk)}{\beta_p-\mu},
\label{eq:I_FF_fit_inc}
\end{eqnarray}
for the other modes.
The total intensity is $I_{\rm NF}(\br)=I_{\rm NF}^{\rm cond}(\br)+I_{\rm NF}^{\rm incoh}(\br)$ ({\it idem} for the FF), with $N=\int I_{\rm NF}(\br) d\br=\int I_{\rm FF}(\bk) d\bk$, and $r=|\br|, k=|\bk|$.
The fact that $I_{\rm NF}^{\rm incoh}$ and $I_{\rm FF}^{\rm incoh}$ only depend on $r$ and $k$ results from a property of the Hermite-Gauss functions.


We performed averages from an ensemble of 2000 realizations of the NF and FF intensity distributions recorded for 
a fixed power $N=7$kW.
In this way, the parameters $(n_0^{eq},\mu)$ have been retrieved by a least square method from the theoretical expressions Eqs.(\ref{eq:I_NF_fit_cond}-\ref{eq:I_FF_fit_inc}).
Figs.~1a-b show examples of the condensate and incoherent contributions to the NF and FF intensity distributions for the same energy $E$ (corresponding to $n_0^{eq}/N \simeq 0.28$).
Note that the averaging over the realizations smooths out the speckle nature of the beam. 
As a result,
the NF and FF representations are equivalent to each other (see Eqs.(\ref{eq:I_NF_fit_cond}-\ref{eq:I_FF_fit_inc})), as evidenced experimentally in Figs.~1a-b.
The validity of the procedure is confirmed by a least square residual error less than 3\% in all cases (2000 realizations) \cite{supplemental}.
Figure~1a-b also report the intensity profiles recorded at the {\it input} of the MMF (`initial condition'), which evidence the power flow toward the fundamental mode.
\\
\\
\\
\\
\\
\\
\\
\\
\\
\\
\\

\begin{widetext}
\begin{center}
\begin{figure}
\includegraphics[width=1\columnwidth]{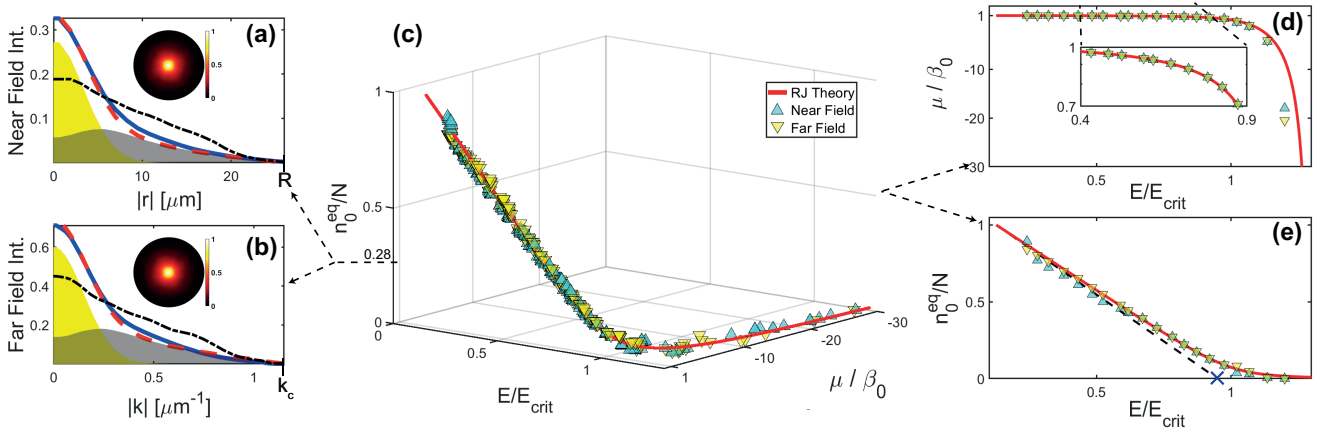}
\caption{
\baselineskip 10pt
{\bf Rayleigh-Jeans condensation:}
Experimental profiles from the NF (a), and FF (b), intensity distributions  recorded at the output of the MMF for the same energy $E/E_{\rm crit} \simeq 0.76$ ($n_0^{eq}/N \simeq 0.28$) with an average over the realizations (solid blue lines).
Theoretical RJ intensity distributions showing the condensate contribution from $p=0$: $I_{\rm NF}^{\rm cond}(r), I_{\rm FF}^{\rm cond}(k)$ [yellow region, from Eq.(\ref{eq:I_NF_fit_cond})], and the incoherent contribution from $p \neq 0$: $I_{\rm NF}^{\rm incoh}(r), I_{\rm FF}^{\rm incoh}(k)$ [grey region, from Eqs.(\ref{eq:I_NF_fit_inc}-\ref{eq:I_FF_fit_inc})], and their respective sums (dashed red lines). 
The insets show the corresponding 2D experimental intensity distributions (averaged over the realizations).
The dashed black lines report the intensity profiles recorded at the input of the MMF (`initial condition').
(c) Condensate fraction $n_0^{eq}/N$ vs ($E/E_{\rm crit}, \mu/\beta_0$): 
The blue (yellow) triangles report the experimental results from the NF (FF) intensity distributions.
The red lines in (c,d,e) report the RJ theory for the MMF used in the experiment.
(d) Corresponding projection $\mu/\beta_0$ vs $E/E_{\rm crit}$: By decreasing the energy $E$ below the critical value $E_{\rm crit}$, the chemical potential $\mu \to  \beta_0^{-}$ 
(inset shows a zoom), which leads to the 
macroscopic population of the fundamental mode, see $n_0^{eq}/N$ vs $E/E_{\rm crit}$ (e).
In (c) 2000 experimental measurements are reported, while an average over the realizations is reported in (d)-(e).
$R=26\mu$m is the fiber radius (a), $k_c=1.15 \mu$m$^{-1}$ the frequency cut-off (b).
The dashed black line in (e) is the condensate fraction in the thermodynamic limit (Eq.(\ref{eq:n_0_TL})) and the blue cross denotes $E_{\rm crit}^*/E_{\rm crit}\simeq 0.95$: The experiment is `close' to the thermodynamic limit.
}
\end{figure}
\end{center}
\end{widetext}
\pagebreak
\newpage
\bigskip
\bigskip
\bigskip
\bigskip
\bigskip
\bigskip
{\it Rayleigh-Jeans condensation.-}
At variance with homogeneous condensation in a bulk medium 
($V(\br) = 0$) \cite{PRL05,nazarenko11},
the presence of a parabolic trapping potential reestablishes condensation in the `thermodynamic limit' in 2D.
There exists a (non-vanishing) critical energy $E_{\rm crit}^*=NV_0/2$ such that $\mu=\beta_0$ \cite{PRA11b,supplemental}.
At this critical point the denominator of the RJ distribution {\it vanishes exactly} for the fundamental mode \cite{PRL05}. 
The singularity is regularized by the macroscopic population of this fundamental mode \cite{PRA11b,supplemental}:
\begin{eqnarray}
n_0^{eq}/N=1-(E-E_0)/(E_{\rm crit}^*-E_0).
\label{eq:n_0_TL}
\end{eqnarray}
Then $n_0^{eq}$ vanishes at $E_{\rm crit}^*$, and $n_0^{eq}/N \to 1$ as the energy decreases to the minimum value $E_0=N\beta_0$.
The phase transition can be {\it equivalently} expressed in term of the temperature, $n_0^{eq}/N=1-T/T_{\rm crit}^*$ with  $T_{\rm crit}^*=N \beta_0^2/V_0$ \cite{PRL05,PRA11b}.
This mechanism of condensation is completely analogous to the quantum Bose-Einstein transition, which originates in the singularity of the Bose distribution when the chemical potential reaches the lowest energy level \cite{stringari,PRL05}.

Because of finite size effects, the experiment does not occur in the strict thermodynamic limit.
The theory of RJ condensation accounting for finite size effects gives the critical energy $E_{\rm crit} = E_0 \big(1+(M-1)/\varrho \big)$, where $\varrho=\sum_{p \neq 0} (p_x+p_y)^{-1}$ \cite{PRA11b,PR14,supplemental}.
Considering the experimental parameters we obtain $E_{\rm crit}^*/E_{\rm crit} \simeq 0.95$ (see the blue cross in Fig.~1e), so that our experiment is relatively close to the thermodynamic limit.

We report in Fig.~1c the condensate fraction $n_0^{eq}/N$ and chemical potential $\mu$ 
as a function of the energy $E$ from 2000 experimental realizations of the NF and FF intensity distributions. 
The corresponding projections $\mu(E)$ and $n_0^{eq}(E)$ are reported in Fig.~1d-e, together with the experimental results averaged over the realizations.
The red line reports the theory of RJ condensation accounting for finite size effects (i.e., beyond the thermodynamic limit) for the MMF used in the experiment, see Refs.\cite{PRA11b,supplemental}. 
By decreasing the energy $E$, the chemical potential $\mu$ increases. 
The transition to condensation takes place when 
$\mu \to \beta_0^{-}$ for $E=E_{\rm crit}$, see Fig.~1d.
Below the transition ($E \le E_{\rm crit}$) the fundamental mode gets macroscopically populated, as shown in Fig.~1e.
The experimental results in Fig.~1d-e are in quantitative agreement with the RJ theory, i.e., $\beta_0$ and $M$ in the theory are fixed by the MMF used in the experiment \cite{supplemental}.
Furthermore, the experimental results are close to the 
thermodynamic limit given by Eq.(\ref{eq:n_0_TL}), see the dashed black line in Fig.~1e.
Note that, as usual, finite size effects make the transition to condensation  `smoother' (red line in Fig.~1e).

\begin{figure}[]
\begin{center}
\includegraphics[width=1\columnwidth]{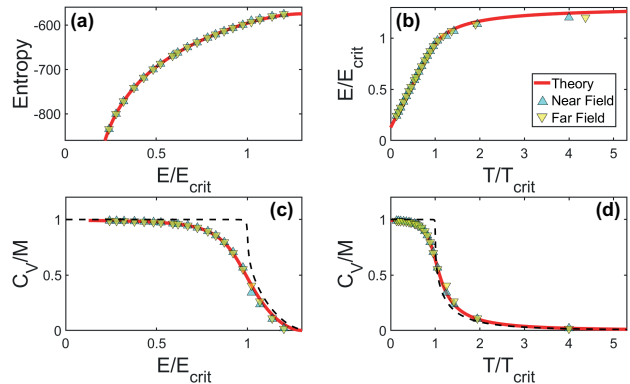}
\caption{\textbf{Thermodynamics of classical condensation:}
(a) Equilibrium entropy $S^{eq}$ vs $E/E_{\rm crit}$;
(b) $E/E_{\rm crit}$ vs $T/T_{\rm crit}$.
Heat capacity per mode $C_V/M$ vs $E/E_{\rm crit}$ (c), and vs $T/T_{\rm crit}$ (d).
The blue (yellow) triangles report the experimental results from the NF (FF) intensity distributions (averaged over the realizations).
The solid red lines report the RJ equilibrium theory for the MMF used in the experiment without adjustable parameters ($M=120$ modes).
For $E<E_{\rm crit}$ (or $T < T_{\rm crit}$) the system exhibits {\it energy equipartition} among the modes and $C_V/M \to 1$, whereas for $E>E_{\rm crit}$ (or $T > T_{\rm crit}$) the {\it equipartition of power} among the modes entails $C_V \to 0$.
By increasing the number of modes, $C_V/M$ exhibits a cusp at the transition to condensation at $E=E_{\rm crit}$ (or $T=T_{\rm crit}$), see the dashed black lines in (c)-(d) for $M=500500$ \cite{supplemental}.
}
\end{center}
\end{figure}

{\it Thermodynamics of classical condensation.-}
We start the thermodynamic study from the {\it equilibrium} entropy $S^{eq}=\sum_p \log(n_p^{eq})-M \log N$, which can be written
\begin{eqnarray}
{S}^{eq}(E)= - \sum_p\log ( \beta_p-\mu(E) ) -M \log\Big( \sum_m \frac{1}{\beta_m-\mu(E)} \Big).
\nonumber
\end{eqnarray}
Figure~2a reports ${S}^{eq}$ vs $E$ by using the experimental data $\mu(E)$ in Fig.~1d.
Note the concavity of the entropy with respect to the energy, 
as required by a self-consistent thermodynamic theory \cite{christodoulides19}.
The heat capacity $C_V=(\partial E/\partial T)_{N,M}$ is known as an important quantity characterizing the quantum BEC transition \cite{stringari}.
In our experiment the transition to condensation is studied by varying $E$ holding fixed $N=$const and $M=$const ($M$ playing a role analogous to the system volume $V$ \cite{christodoulides19}).
Making use of the energy-temperature relation $E$ vs $T=(\partial E/ \partial S^{eq})_{N,M}$ in Fig.~2b, we obtain \cite{supplemental}:
\begin{eqnarray}
C_V(E)=M - \frac{\big(  \sum_p (\beta_p-\mu(E))^{-1} \big)^2}{\sum_p (\beta_p -\mu(E))^{-2}}.
\label{eq:c_v_expl}
\end{eqnarray}
The heat capacity exhibits distinguished properties in comparison with quantum BECs \cite{stringari}.
Below the transition ($E < E_{\rm crit}$) we have $\mu \to \beta_0^-$. 
Then writing the energy in the form $E=T M+\beta_0 N$, it becomes apparent that the heat capacity is given by the number of degrees of freedom $C_V =M$, as expected from the {\it theorem of energy equipartition} inherent to classical statistical mechanics, see Fig.~2c-d.
This contrasts with the quantum notion of `frozen degrees of freedom', together with the low temperature limit $C_V \to 0$ inherent to a quantum gas.

Far above the BEC transition a quantum gas behaves as a classical gas featured by a constant heat capacity $C_V(T)=$const \cite{stringari}.
At variance with a {\it classical gas}, we observe in our classical wave system that $C_V \to 0$ for $E > E_{\rm crit}$ (or $T > T_{\rm crit}=N\beta_0/\varrho$), see Fig.~2c-d.
Actually, the equilibrium properties of waves are of different nature than those of a gas:
Above the transition to condensation the equilibrium state no longer exhibits energy equipartition among the modes, but instead a {\it modal equipartition of the power}, viz $n_p^{eq} \sim T /(-\mu)$  for $-\mu \gg \beta_p$.
This state is the most disordered among all equilibrium states with $S^{eq}_{\rm max}=-M \log M$ for $E = E_m \simeq \frac{2}{3} N V_0$ and $1/T=(\partial S/\partial E)_{N,M} \to 0$.
This means that the equilibrium state is {\it not constrained by the conservation of the energy} $E$ (the Lagrange multiplier $1/T$ is zero), but solely by the conservation of the power $N$, which merely explains the equipartition of power among the modes \cite{PRX17}.
Accordingly, a variation of the temperature $T$ does not affect $E$, which entails $C_V \to 0$ as $E \to E_m$.
Approaching the continuous thermodynamic limit, $C_V$ exhibits a cusp featured by an infinite derivative  at the critical point $E=E_{\rm crit}$ (or $T=T_{\rm crit}$), see the dashed lines in Fig.~2c-d.
Note that energies $E>E_m$ could be excited with artificial initial conditions, a feature that will be considered in future works \cite{christodoulides19,supplemental}.


{\it Conclusion and perspectives.-} 
We have reported the  experimental observation of the equilibrium condensation of classical optical waves in quantitative agreement with the RJ theory. 
Far above the transition we have shown that the field exhibits an equipartition of power among the modes 
and a vanishing heat capacity ($C_V(T) \to 0$). 
By decreasing the energy (or temperature), a transition occurs from the normal state toward the condensed state, which is featured by a constant heat capacity with a macroscopic population of the fundamental mode and an energy equipartition among the other modes.


Our experiments in MMFs pave the way for a {\it thermodynamic control of the coherence properties of light} \cite{christodoulides19}.
For instance, a thermalized speckle beam in the normal state 
can be adiabatically cooled toward the condensed state owing to a potential sink in a manufactured optical fiber, in complete analogy with the adiabatic formation of quantum BECs \cite{stringari}.


{\it Acknowledgements.-} 
The authors are grateful to C. Michel, A. Tonello, P. B\'ejot, S. Gu\'erin, V. Couderc, and A. Barth\'elemy for fruitful discussions.
We acknowledge financial support from the French ANR under Grant No. ANR-19-CE46-0007 (project ICCI),
iXcore research foundation, EIPHI Graduate School (Contract No. ANR-17-EURE-0002), French program ``Investissement d'Avenir," Project No. ISITE-BFC-299 (ANR-15 IDEX-0003);  H2020 Marie Sklodowska-Curie Actions (MSCA-COFUND) (MULTIPLY Project No. 713694). 

\end{document}